\documentclass[aps,prd,reprint,onecolumn,10pt,eqsecnum,showpacs,nofootinbib,amsfonts,amssymb,amsmath,longbibliography,superscriptaddress,floatfix]{revtex4-1}
\usepackage[colorlinks=true,allcolors=blue, pdfusetitle]{hyperref}
\usepackage{bm}
\usepackage{mathrsfs}
\usepackage[dvipsnames]{xcolor}
\usepackage{mathtools}
\usepackage{graphicx}
\usepackage{orcidlink}
\usepackage{tensor}

\usepackage{algorithm}
\usepackage{algpseudocode}

\usepackage[commentmarkup=uwave]{changes}
\definechangesauthor[name=hs, color=cyan]{hs}

\usepackage{fontawesome5}

\usepackage{ragged2e}

\usepackage{tikz}

\usetikzlibrary{shapes.geometric, arrows,positioning}

\tikzstyle{startstop} = [rectangle, rounded corners, minimum width=3cm, minimum height=1cm, text centered, draw=black, fill=orange!30]
\tikzstyle{process} = [rectangle,rounded corners, minimum width=3cm, minimum height=1cm, text centered, draw=black, fill=blue!15]
\tikzstyle{decision} = [rectangle,rounded corners, minimum width=3cm, minimum height=1cm, text centered, draw=black, fill=green!20]
\tikzstyle{arrow} = [->, >=stealth]

\graphicspath{{../plots/}}

\usepackage[T1]{fontenc}
\usepackage{lmodern}
\usepackage[utf8]{inputenc}
\usepackage{microtype}

\DeclareMathAlphabet{\mathbfsf}{\encodingdefault}{\sfdefault}{bx}{sl}

\usepackage[normalem]{ulem}

\begin{document}
	
\title{Thermodynamic phase transition and Joule-Thomson expansion of a quantum corrected black hole in AdS spacetime}

\author{Rui-Bo Wang\,\orcidlink{0009-0003-3198-3310}}
\email[Rui-Bo Wang: ]{wangrb2021@lzu.edu.cn}
\affiliation{School of Physical Science and Technology, Lanzhou University, Lanzhou, Gansu 730000, China}

\author{Lei You\,\orcidlink{0009-0004-5900-3318}}
\email[Lei You: ]{youl21@lzu.edu.cn}
\affiliation{School of Physical Science and Technology, Lanzhou University, Lanzhou, Gansu 730000, China}

\author{Shi-Jie Ma\,\orcidlink{0000-0003-4423-0142}}
\email[Shi-Jie Ma: ]{220220939731@lzu.edu.cn}
\affiliation{School of Physical Science and Technology, Lanzhou University, Lanzhou, Gansu 730000, China}

\author{Jian-Bo Deng\,\orcidlink{0000-0002-0586-6220}}
\email[Jian-Bo Deng(corresponding author): ]{dengjb@lzu.edu.cn}
\affiliation{School of Physical Science and Technology, Lanzhou University, Lanzhou, Gansu 730000, China}

\author{Xian-Ru Hu\,\orcidlink{0000-0001-9818-8635}}
\email[Xian-Ru Hu: ]{huxianru@lzu.edu.cn}
\affiliation{School of Physical Science and Technology, Lanzhou University, Lanzhou, Gansu 730000, China}

\date{\today}

\begin{abstract}
The thermodynamics in the extended phase space of a quantum corrected black hole (BH) proposed recently is presented in this work. Our study shows that the phase transition behavior of the BH is analogous to that of conventional Schwarzschild BH in anti-de Sitter (AdS) space; however, a critical temperature exists such that when the BH temperature exceeds this critical value, the small BH phase and the large BH phase become separated, and no phase transition occurs. Due to the introduction of the quantum parameter $\xi$, the BH equation of state splits into two branches. One branch reduces to the Schwarzschild-AdS case as $\xi\to0$, with its phase transition pressure lower than the critical pressure; another branch's phase transition pressure is greater than the critical pressure. The study shows that the $T-r_{+}$ phase transition and heat capacity are similar to those of the Schwarzschild-AdS BH. The Joule-Thomson expansion is divided into two stages: in the earlier stage, the BH pressure increases until it reaches a maximum; in the later stage, the pressure gradually decreases. In each stage, the BH may undergo an inversion point, resulting in the inversion curve with two branches. In addition, each stage has a minimum inversion mass, below which any BH (in each respective stage) has no inversion point.

\hspace*{\fill}


\end{abstract}

\maketitle
\section{Introduction}\label{Sect1}
Due to the inconsistency between general relativity (GR) and quantum theory, as well as issues such as singularities~\cite{GR1,GR2,GR3,GR4,GR5}, quantum gravity has long been a subject of extensive interest. One approach to achieve spacetime quantization is non-commutative geometry~\cite{NC1,NC2,NC3,NC4,NC5,NC6,NC7,NC8}, which is characterized by the commutation relation among spacetime coordinate operators $[\hat{x^{\mu}},\hat{x^{\nu}}]=i\theta^{\mu\nu}$, where $\theta^{\mu\nu}$ is a anti-symmetric constant matrix. A. Smailagic's research shows that noncommutative effects eliminated the point-like mass distribution~\cite{GR2,GR3}, a phenomenon that may circumvent the BH singularity problem. Inspired by this, P. Nicolini and K. Nozari respectively proposed that the point mass distribution model (represented by the Dirac delta function) can be replaced by either a Gaussian distribution or a Lorentzian distribution~\cite{NCLorentz,NCGauss}. In addition, effective field theory (EFT) and loop quantum gravity (LQG) are also candidate theories for quantum gravity. In Ref.~\cite{EFT1} the authors analyze low-energy one-loop quantum corrections to the Schwarzschild geometry. In their approach, GR is treated as a quantum field theory (QFT) within the framework of EFT, implying that at low energies the degrees of freedom organize themselves as quantum fields governed by a local Lagrangian~\cite{EFT2}. Similarly, LQG has received significant attention~\cite{LQG1,LQG2,LQG3,LQG4,LQG5,LQG6,LQG7,LQG8}. In LQG, the minimal fundamental units of spacetime are suggested as a series of discrete, elementary loops. One interesting result is that by studying the gravitational collapse of spherically symmetric dust matter in LQG, researchers have constructed a quantum corrected Schwarzschild BH and found that it can be interpreted as the formation of a white hole~\cite{LQGf,LQGshadow1,LQGshadow2}.

The BH discussed in this paper was proposed in a recent study that aimed to address the longstanding issue of general covariance in quantum gravity models. It is suggested that the classical GR could be modified by Hamiltonian constraint formulation~\cite{Ham1,Ham2,Ham3}. In this approach, an issue arose: how to keep the general diffeomorphism covariance of spacetime theory~\cite{xioptics}. Recently, this problem was effectively addressed in the framework of spherically symmetric vacuum gravity~\cite{sourcetheory}. In the study of Zhang \textit{et al}.~\cite{sourcetheory}, general covariance is rigorously formulated into a set of constraint equations. Based on this formulation, they derived the equations of motion for the effective Hamiltonian constraint, which depends on a quantum parameter $\xi$. Furthermore, they provided two candidate forms for the effective Hamiltonian constraint, each corresponding to a quantum corrected Schwarzschild spacetime. In this work, we investigate the thermodynamics of one such static, spherically symmetric quantum corrected BH.

The thermodynamics of AdS BHs in the extended phase space is a novel and highly intriguing subject. Within this framework, the cosmological constant term in Einstein equation is interpreted as the energy-momentum tensor of a special ''vacuum static ideal fluid'', and it is assumed that the BH achieves phase equilibrium with the vacuum fluid, thereby acquiring pressure~\cite{kastor}. To ensure that the BH exhibits a positive pressure, the cosmological constant is required to be negative, implying that the BH resides in an AdS spacetime. Consequently, the AdS BH is effectively regarded as a unique thermodynamic system, which has attracted extensive discussion in the academic community. In 1983, Hawking first investigated the thermodynamics of Schwarzschild-AdS BHs, revealing the existence of phase transitions within these systems~\cite{Hawking}. This pioneering work is widely regarded as the inception of the field of AdS BH thermodynamics. Numerous studies have explored the thermodynamics of AdS BHs~\cite{thermo1,thermo2,thermo3,thermo4,thermo5,thermo6,thermo7,thermo8,thermo9,thermo10,thermo11,thermo12,thermo13,thermo14}. The thermodynamic properties of BHs are influenced by several additional parameters, such as electric charge~\cite{charged1,charged2,pvcri,JTcharged}, nonlinear magnetic charge~\cite{BD1,BD2,BD3}, dark matter~\cite{DM1,DM2}, nonlinear electrodynamics models~\cite{NED1,NED2,NED3,NED4,NED5,NED6,NED7}, noncommutative geometry~\cite{NCthermo1,NCthermo2,NCthermo3,NCthermo4}, lower or higher-dimensional gravity theories~\cite{dimension1,dimension2,dimension3,dimension4}, and various other modified gravity models~\cite{mGR1}. Moreover, as a candidate theory for reconciling gravity and quantum physics, the AdS/CFT correspondence also offers a novel perspective on the thermodynamics of BHs in AdS spacetime~\cite{adscft1,adscft2}. Many BHs exhibit thermodynamic properties analogous to those of Van der Waals system, implying a profound connection between gravitational theory, quantum physics and thermodynamics, and thereby revealing a richer array of physical properties for these enigmatic objects.

The structure of this paper is organized as follows. In Sect.~\ref{Sect2}, we introduce the quantum corrected BH that is the focus of this paper. Under the assumption that this quantum gravity decouples from the cosmological constant, we derive the BH metric in AdS spacetime and compute its thermodynamic functions as well as the corrected first law. Sect.~\ref{Sect3} will provide a detailed investigation into the thermodynamic phase transitions and critical phenomena of the BH. The isobaric heat capacity and the Gibbs free energy of the BH are researched in Sect.~\ref{Sect4} and Sect.~\ref{Sect5} respectively. In Sect.~\ref{Sect6}, we conduct a detailed investigation of the constant mass expansion of the BH and present its inversion curve. Finally, the summary and outlook of this work will be presented in Sect.~\ref{Sect7}. Our study will demonstrate the influence of the quantum correction parameter on the thermodynamics of conventional Schwarzschild-AdS BH, thereby enriching the fruits of research on quantum BH thermodynamics and providing a valuable reference for subsequent investigations.

For computational convenience, natural units $G=\hbar=c=k_{B}=1$ are adopted throughout this work.

\section{Thermodynamic functions and the first law}\label{Sect2}
\begin{figure}[htbp]
	\centering
	\includegraphics[width=0.56\textwidth]{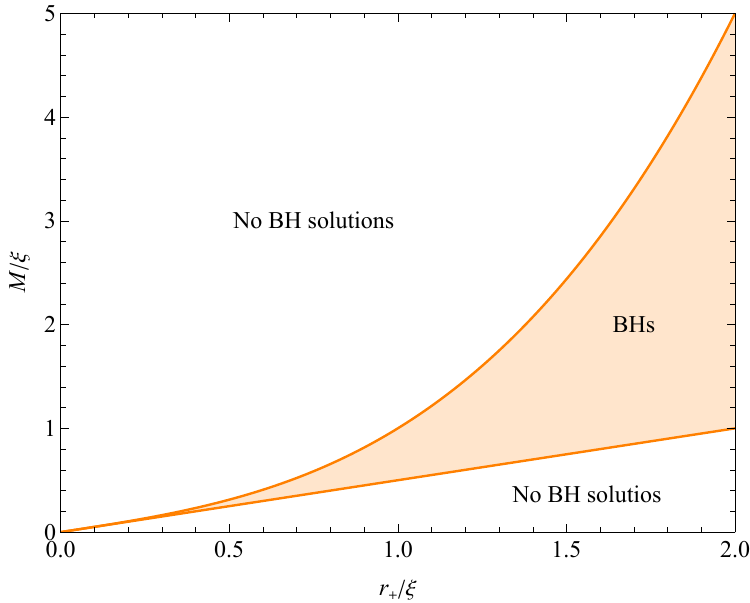}
	\caption{Condition of BH's existence in $\left(r_{+},M\right)$ coordinate. The orange region signifies the parameter range in which BH solutions exist, while the remaining areas indicate the absence of BH solutions.}\label{BHcd}
\end{figure}

The BH considered in this paper is a quantum corrected BH, described by a static spherically symmetric spacetime~\cite{sourcetheory}
\begin{equation}\label{eq2_1}
	ds^{2}=-f\left(r\right)dt^{2}+\frac{1}{f\left(r\right)}dr^{2}+r^{2}d\theta^{2}+r^{2}\sin^{2}\theta d\phi^{2},
\end{equation}
with $f\left(r\right)$ being
\begin{equation}\label{eq2_2}
	f\left(r\right)=1-\frac{2M}{r}\left(1+\frac{\xi^{2}}{r^{2}}\left(1-\frac{2M}{r}\right)\right),
\end{equation}
where $\xi$ is a quantum parameter, which is considered to be on the order of the Planck length $\ell_{p}$. This BH corresponds to an equivalent energy distribution
\begin{equation}\label{eq2_3}
	T_{t\left(\mathrm{BH}\right)}^{t}=-\frac{\xi^{2}\left(r-2M\right)\left(r-6M\right)}{8\pi r^{6}}.
\end{equation}
Assuming that this quantum gravity decouples from the cosmological constant $\Lambda$, the Einstein equation with $\Lambda$ is
\begin{equation}\label{eq2_4}
	R_{\mu}^{\nu}-\frac{1}{2}\delta_{\mu}^{\nu}R=8\pi\left(T_{\mu\left(\mathrm{BH}\right)}^{\nu}+T_{\mu\left(\Lambda\right)}^{\nu}\right),
\end{equation}
where $T_{\mu\left(\Lambda\right)}^{\nu}=-\frac{\Lambda}{8\pi}\delta_{\mu}^{\nu}$ represents an equivalent energy-momentum tensor corresponding to the contribution of the cosmological constant term. Thus the metric of this quantum corrected BH in AdS spacetime is obtained as
\begin{equation}\label{eq2_5}
	f\left(r\right)=1-\frac{2M}{r}\left(1+\frac{\xi^{2}}{r^{2}}\left(1-\frac{2M}{r}\right)\right)-\frac{\Lambda r^{2}}{3}.
\end{equation}
The BH solution satisfies $f\left(r_{+}\right)=0$, where $r_{+}$ is the event horizon radius. The mass of BH could be solved
\begin{equation}\label{eq2_6}
	M=\frac{2r_{+}\xi^{2}+r_{+}^{3}\left(1\pm\sqrt{\alpha}\right)}{4\xi^{2}},
\end{equation}
where
\begin{equation}\label{eq2_7}
	\alpha=1+\frac{4}{3}\Lambda\xi^{2}.
\end{equation}
Here, we discard the positive-sign branch because this branch does not reduce to the Schwarzschild-AdS case as $\xi\to0$, but instead diverges to infinity. It is required that $\alpha\geqslant0$, which gives
\begin{equation}\label{eq2_8}
	\Lambda\geqslant\Lambda_{min}=-\frac{3}{4\xi^{2}}.
\end{equation}
On the other hand, the existence of BH requires $M>0, r_{+}>0, \Lambda<0$ (AdS spacetime), which reduces this parameter constraint
\begin{equation}\label{eq2_9}
	\frac{r_{+}}{2}<M<\frac{1}{2}\left(r_{+}+\frac{r_{+}^{3}}{\xi^{2}}\right),
\end{equation}
or
\begin{equation}\label{eq2_10}
	\frac{-3^{\frac{2}{3}}\xi^{\frac{4}{3}}+\xi^{\frac{2}{3}}3^{\frac{1}{3}}\beta^{\frac{2}{3}}}
	{3\beta^{\frac{1}{3}}}<r_{+}<2M.
\end{equation}
where $\beta=9M+\sqrt{81M^{2}+3\xi^{2}}$. The condition of BH's existence is shown in Fig.~\ref{BHcd}.

The BH's temperature is defined by its surface gravity
\begin{equation}\label{eq2_11}
	T=\frac{\partial_{r}f}{4\pi}\bigg|_{r_{+}}=\frac{\sqrt{\alpha}}{4\pi r_{+}}+\frac{3r_{+}\left(\sqrt{\alpha}-\alpha\right)}{8\pi\xi^2}
\end{equation}
In the extended phase space, the energy-momentum tensor $T_{\mu\left(\Lambda\right)}^{\nu}$ contributed by the cosmological constant is regarded as an equivalent ''vacuum static ideal fluid''. It is assumed that the BH reaches a phase equilibrium with this fluid, acquiring its pressure. The pressure of the BH is
\begin{equation}\label{eq2_12}
	P=-\frac{\Lambda}{8\pi}.
\end{equation}
Eq.~\ref{eq2_8} gives the maximum pressure of BH
\begin{equation}\label{eq2_13}
	P_{max}=-\frac{\Lambda_{min}}{8\pi}=\frac{3}{32\pi\xi^{2}}.
\end{equation}
According to the studies in references, when the energy-momentum tensor outside the BH event horizon explicitly includes the BH's mass $M$, the first law of BH thermodynamics takes a corrected form~\cite{FL1,FL2,FL3,BD2}
\begin{equation}\label{eq2_14}
	WdM=TdS+VdP,
\end{equation}
where correction function $W$ is
\begin{equation}\label{eq2_15}
	W=1+\int_{r_{+}}^{\infty}4\pi r^{2}\frac{\partial T_{t}^{t}}{\partial M}dr=\sqrt{\alpha}.
\end{equation}
The entropy of the BH is the Bekenstein-Hawking function~\cite{HB1,HB2}
\begin{equation}\label{eq2_16}
	S=\int\frac{W}{T}dM=\int\frac{W}{T}\frac{\partial M}{\partial r_{+}}dr_{+}=\pi r_{+}^{2}.
\end{equation}
The BH's thermodynamic volume is
\begin{equation}\label{eq2_17}
	V=W\left(\frac{\partial M}{\partial P}\right)_{S}=\frac{4\pi r_{+}^{3}}{3}.
\end{equation}

\section{Thermodynamic phase transition}\label{Sect3}
\begin{figure}[htbp]
	\centering
	\begin{minipage}{0.495\textwidth}
	\centering
	\includegraphics[width=\linewidth]{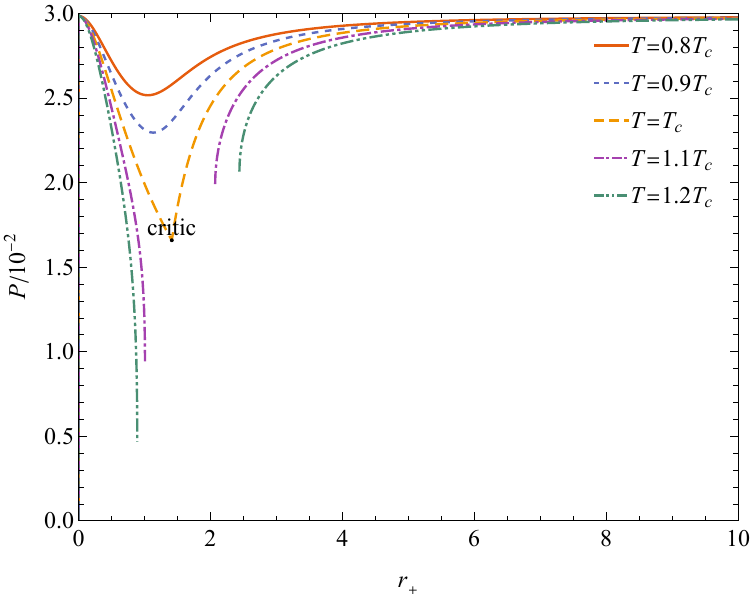}
	\caption{BH's $P_{+}-r_{+}$ diagram. Black point ''critic'' is the critical point. We set $\xi=1$.}\label{Prp}
	\end{minipage}
	\hfill
	\begin{minipage}{0.495\textwidth}
	\centering
	\includegraphics[width=\linewidth]{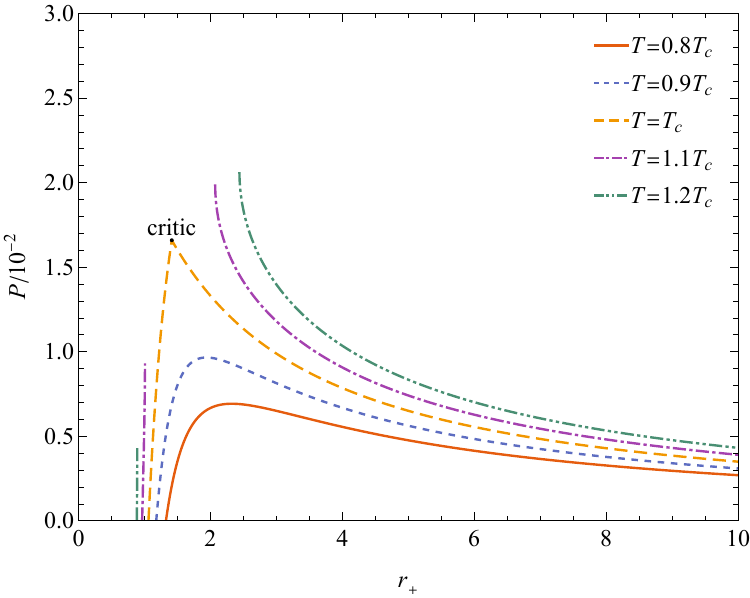}
	\caption{BH's $P_{-}-r_{+}$ diagram. Black point ''critic'' is the critical point. We set $\xi=1$.}\label{Prn}
	\end{minipage}
\end{figure}

The equation of state of BH is solved from Eq.~\ref{eq2_11}
\begin{equation}\label{eq3_1}
	P_{\pm}=\frac{T}{4r_{+}}-\frac{1}{16\pi r_{+}^{2}}+\frac{3}{64\pi\xi^{2}}-\frac{\xi^{2}}{48\pi r_{+}^{4}}\pm\left(\frac{1}{3r_{+}^{2}}+\frac{1}{2\xi^{2}}\right)\frac{\sqrt{9r_{+}^{4}+12r_{+}^{2}\left(1-8\pi r_{+}T\right)\xi^{2}+4\xi^{4}}}{32\pi r_{+}^{2}}.
\end{equation}
It is evident that the $P-r_{+}$ diagram of the BH consists of two branches, corresponding respectively to the positive-sign branch $P_{+}$ (Fig.~\ref{Prp}) and negative-sign branch $P_{-}$ in the above function (Fig.~\ref{Prn}).

In particular, there exists a critical temperature 
\begin{equation}\label{eq3_2}
 	T_{c}=\frac{1}{3\sqrt{2}\pi\xi}.
\end{equation}
When $T>T_{c}$, the small BH phase ($r_{+}<r_{1}\left(T\right)$) and the large BH phase ($r_{+}>r_{2}\left(T\right)$) become distinctly separated, and no phase transition occurs. $r_{1}\left(T\right)$ and $r_{2}\left(T\right)$ are the smaller and larger positive real roots of equation with parameter $T$ 
\begin{equation}\label{eq3_3}
	9r_{+}^{4}+12r_{+}^{2}\xi^{2}-96\pi r_{+}^{3}T\xi^{2}+4\xi^{4}=0,
\end{equation}
respectively.

When $T=T_{c}$, the small BH-large BH phase transition occurs at the critical radius
\begin{equation}\label{eq3_4}
	r_{+c}=\sqrt{2}\xi.
\end{equation}
At the critical point, BH reaches its critical pressure 
\begin{equation}\label{eq3_5}
	P_{c}=\frac{5}{96\pi\xi^{2}}=\frac{5}{9}P_{max}.
\end{equation}
It is important to note that, in this case, the critical point cannot be determined using the conventional critical point condition $\partial P/\partial r_{+}=\partial^{2} P/\partial r_{+}^{2}=0$. As can be seen directly from Fig.~\ref{Prp} and Fig.~\ref{Prn}, the critical behavior of this type of BH is entirely different from that of a Van der Waals fluid. At the critical point, $dP/dr_{+}\big|_{r_{+c}^{-}}\neq dP/dr_{+}\big|_{r_{+c}^{+}}$, implying that $dP/dr_{+}$ does not exist. Nevertheless, only when $T<T_{c}$, there exists a solution of $dP/dr_{+}=0$, which corresponds to the phase transition.

\begin{figure}[htbp]
	\centering
	\includegraphics[width=0.56\textwidth]{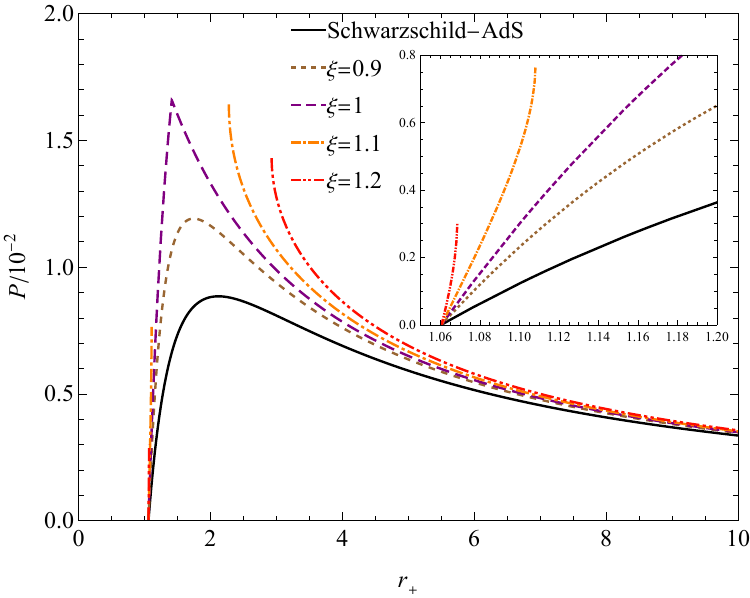}
	\caption{BH's $P_{-}-r_{+}$ diagram for different quantum parameters $\xi$. Black line corresponds to the Schwarzschild-AdS BH ($\xi=0$). We set $T=T_{c}\left(\xi=1\right)$.}\label{PrSch}
\end{figure}

This critical point gives a dimensionless constant
\begin{equation}\label{eq3_6}
	\rho_{c}=\frac{2P_{c}r_{+c}}{T_{c}}=\frac{5}{8}.
\end{equation}
So this quantum corrected BH shows a larger critical ratio than Van der Waals system ($\frac{3}{8}$)~\cite{pvcri}. Furthermore, please notice that the equation of state $P_{-}$ shown in Fig.~\ref{Prn} will reduce to the Schwarzschild-AdS case in the limit $\xi\to0$. In Fig.~\ref{Prn}, during an isothermal expansion process, the pressure of the small BH increases rather than decreases, signifying that it corresponds to an unstable phase. In contrast, the large BH exhibits decreasing pressure with increasing radius. However, the situation in the BH phase diagram presented in Fig.~\ref{Prp} is completely opposite, where the small BH corresponds to the stable phase, while the large BH corresponds to the unstable phase. This characteristic arises due to the quantum parameter $\xi$, and this feature does not occur in the ordinary Schwarzschild-AdS BH. The BH's $P_{-}-r_{+}$ diagram for different quantum parameters $\xi$ is shown in Fig.~\ref{PrSch}.

Specifically, when $P_{-}=0$, the horizon radius of the BH is independent of the size of the quantum parameter $\xi$
\begin{equation}\label{eq3_7}
	r_{+}\big|_{P=0}=\frac{1}{4\pi T},
\end{equation}
which is also the result of Schwarzschild-AdS BH.

\begin{figure}[htbp]
	\centering
	\includegraphics[width=0.56\textwidth]{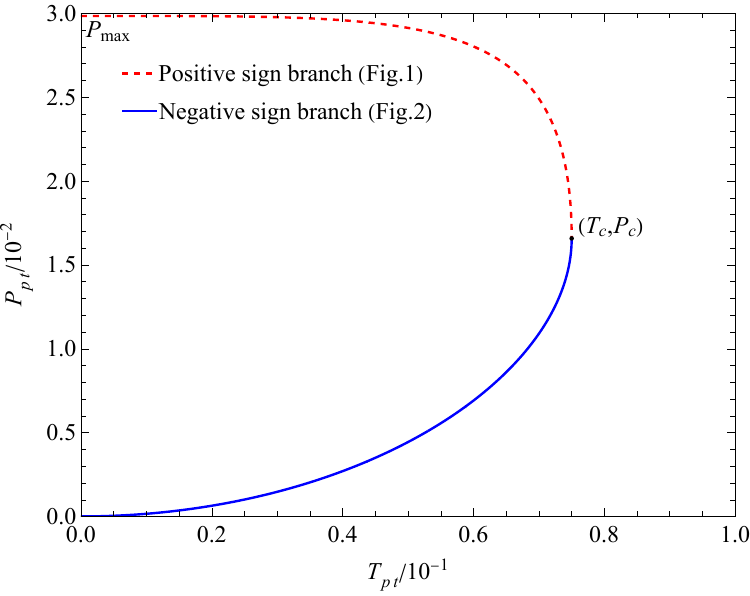}
	\caption{Phase transition points in $T-P$ coordinate. Point $\left(T_{c},P_{c}\right)$ is the critical point. We set $\xi=1$.}\label{ptf}
\end{figure}

The phase transition occurs at
\begin{equation}\label{eq3_8}
	\frac{\partial P}{\partial r_{+}}=0.
\end{equation}
The variation of the BH phase transition pressure $P_{pt}$ with respect to the phase transition temperature $T_{pt}$ is illustrated in Fig.~\ref{ptf}.
Thus, it is apparent that the critical temperature represents the maximum phase transition temperature. Moreover, if we consider only the BH pressure as described in Fig.~\ref{Prn}, then the critical pressure will also correspond to the maximum phase transition pressure. This characteristic is analogous to that observed in the Van der Waals system.

\begin{figure}[htbp]
	\centering
	\begin{minipage}{0.504\textwidth}
	\centering
	\includegraphics[width=\linewidth]{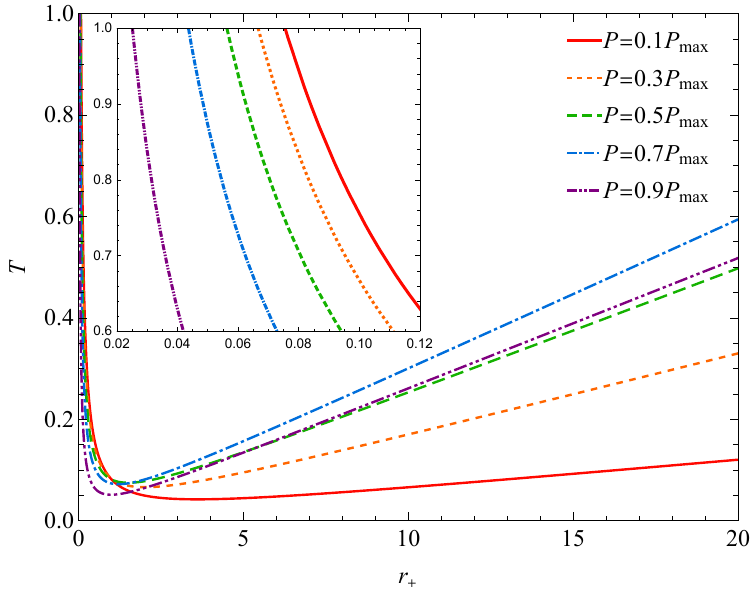}
	\caption{BH's $T-r_{+}$ curve corresponding to the isobaric process. We set $\xi=1$.}\label{Trf}
	\end{minipage}
	\hfill
	\begin{minipage}{0.49\textwidth}
	\centering
	\includegraphics[width=\linewidth]{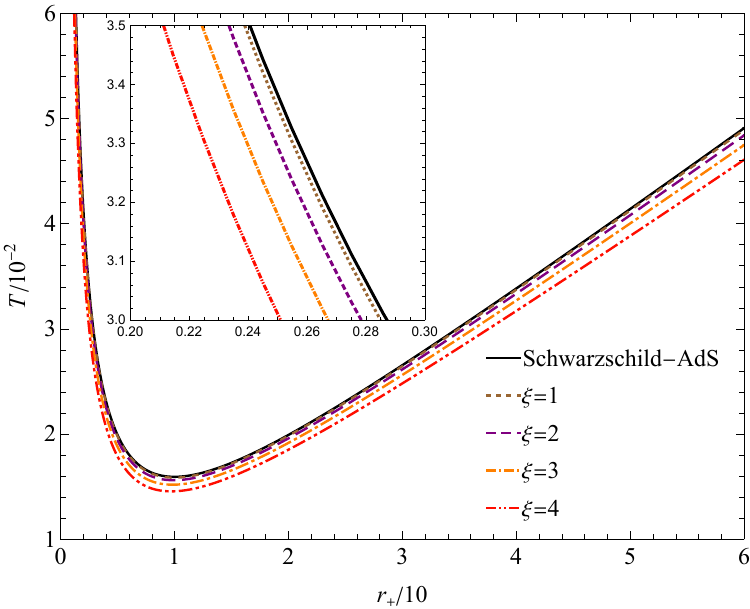}
	\caption{BH's $T-r_{+}$ diagram for different quantum parameters $\xi$. Black line ($\xi=0$) represents the Schwarzschild-AdS BH. We set $\Lambda=-0.01$.}\label{TrSch}
	\end{minipage}
\end{figure}

The BH's $T-r_{+}$ diagram is shown in Fig.~\ref{Trf} and Fig.~\ref{TrSch}. It is evident that the BH exhibits a small BH-large BH phase transition similar to that of the Schwarzschild-AdS BH. The BH pressure initially increases with the horizon radius (corresponding to the small BH phase) and subsequently decreases (corresponding to the large BH phase). This phase transition, in contrast to that observed in Van der Waals fluids, occurs instantaneously at the maximum pressure point.

In particular, phase separation does not occur in BH's $T-r_{+}$ diagram. It can be verified that $T>0$ always holds. When $P=P_{max}$, the BH is an extremal BH, consistently maintaining a zero temperature $T\equiv0$. For any given pressure, the BH temperature always undergoes a small BH-large BH phase transition. The temperature of the small BH decreases as the horizon radius increases, whereas the large BH exhibits the opposite behavior.

The phase transition temperature $T_{pt}$ satisfies
\begin{equation}\label{eq3_9}
	\frac{\partial T}{\partial r_{+}}=0.
\end{equation}
Using this condition, one could prove that at $P=P_{c}$, $T_{pt}$ has its maximum
\begin{equation}\label{eq3_10}
	T_{pt}^{max}=T_{c}.
\end{equation}
This also, from the perspective of the $T-r_{+}$ diagram, demonstrates that BHs with temperatures exceeding the critical temperature cannot undergo phase transitions.

\section{Heat capacity}\label{Sect4}
According to
\begin{equation}\label{eq4_1}
	S=\pi\left(\frac{3V}{4\pi}\right)^{\frac{2}{3}},
\end{equation}
the BH's isohoric heat capacity is zero
\begin{equation}\label{eq4_2}
	C_{V}=T\left(\frac{\partial S}{\partial T}\right)_{V}=0.
\end{equation}
The BH's isobaric heat capacity is
\begin{equation}\label{eq4_3}
	C_{p}=T\left(\frac{\partial S}{\partial T}\right)_{P}=\frac{2\pi r_{+}^{2}\left(24P\pi r_{+}^{4}-\xi^{2}+r_{+}^{2}\sqrt{9-96P\pi\xi^{2}}\right)}{-3r_{+}^{2}+24P\pi r_{+}^{4}+\xi^{2}}.
\end{equation}
\begin{figure}[htbp]
	\centering
	\begin{minipage}{0.49\textwidth}
		\centering
		\includegraphics[width=\linewidth]{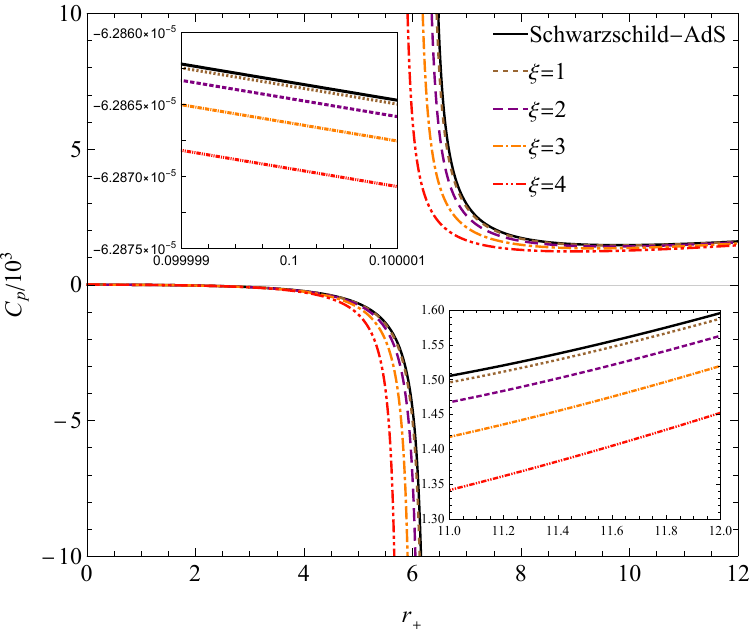}
		\caption{Isobaric heat capacity for different quantum parameters $\xi$. Black line represents the Schwarzschild-AdS BH. We set $P=0.001$.}\label{cpSch}
	\end{minipage}
	\hfill
	\begin{minipage}{0.49\textwidth}
		\centering
		\includegraphics[width=\linewidth]{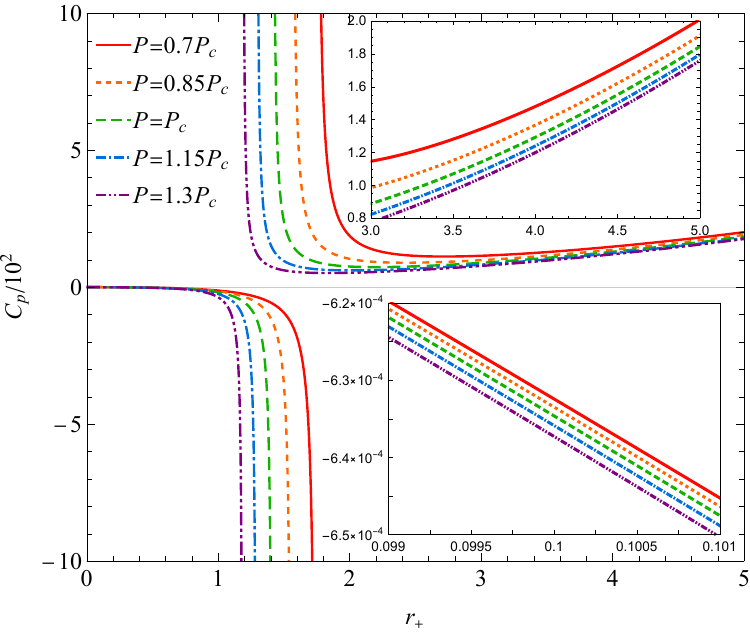}
		\caption{Isobaric heat capacity for different pressure. We set $\xi=1$.}\label{cpf}
	\end{minipage}
\end{figure}

The heat capacity as a function of the horizon radius is presented in Fig.~\ref{cpSch} and Fig.~\ref{cpf}. In comparison with Fig.~\ref{Trf}, it is readily evident that the small BH-satisfying $\partial T/\partial r_{+}<0$-exhibits a negative heat capacity, corresponding to an unstable phase, whereas the large BH-described by $\partial T/\partial r_{+}>0$-displays a positive heat capacity, indicative of a stable phase. At the phase transition point, the heat capacity diverges to infinity. Moreover, when the horizon radius is sufficiently small, the heat capacity exhibits asymptotic behavior
\begin{equation}\label{eq4_4}
	C_{p}=-2\pi r_{+}^{2}+\mathcal{O}\left(r_{+}\right)^{4}.
\end{equation}
In the limit as $r_{+}\to\infty$, the BH's isobaric heat capacity exhibits a quadratic divergence
\begin{equation}\label{eq4_5}
	C_{p}=2\pi r_{+}^{2}+\frac{3+\sqrt{9-96P\pi\xi^{2}}}{12P}+\mathcal{O}\left(r_{+}\right)^{-2}.
\end{equation}

\section{Gibbs free energy}\label{Sect5}
The BH's Gibbs free energy is
\begin{equation}\label{eq5_1}
	G=M-TS=\frac{r_{+}}{2}-4P\pi r_{+}^{3}-\frac{1}{12}r_{+}\sqrt{9-96P\pi\xi^{2}}+\frac{5r_{+}^{3}\left(3-\sqrt{9-96P\pi\xi^{2}}\right)}{24\xi^{2}}.
\end{equation}
\begin{figure}[htbp]
	\centering
	\includegraphics[width=0.56\textwidth]{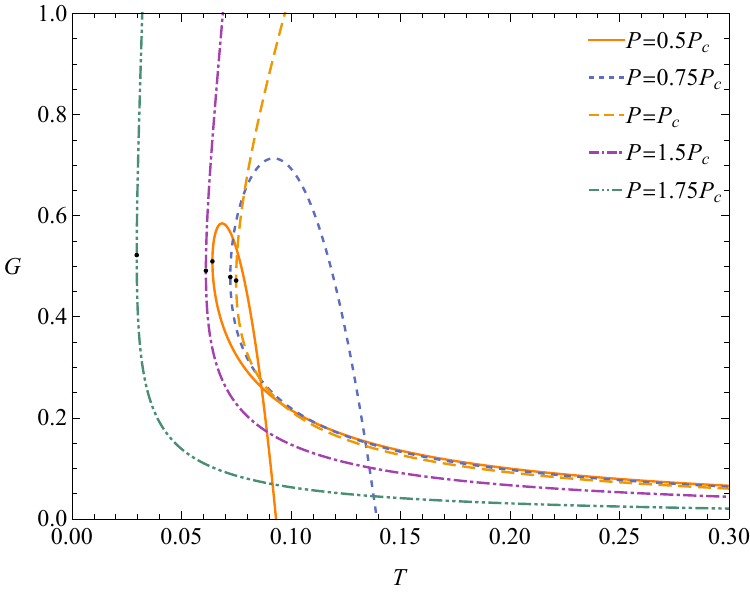}
	\caption{Gibbs free energy of the BH for different pressures. The black points in the figure are phase transition points. We set $\xi=1$.}\label{GTP}
\end{figure}

One could calculate the differential form of $G$
\begin{equation}\label{eq5_2}
	dG=\left(W^{-1}-1\right)TdS-SdT+W^{-1}VdP,
\end{equation}
which gives
\begin{equation}\label{eq5_3}
	\left(\frac{\partial G}{\partial T}\right)_{P}=-S+\left(W^{-1}-1\right)C_{p}.
\end{equation}
Phase transition satisfies $C_{p}=\infty$, so in the BH's $G-T$ diagram, the points at which $\partial G/\partial T=\infty$ are exactly the phase transition points.

Specifically, when $P<P_{c}$, there exists a point at which
\begin{equation}\label{eq5_4}
	\frac{\partial G}{\partial T}=0,
\end{equation}
shown as orange line ($P=0.5P_{c}$) and lightblue line ($P=0.75P_{c}$) in Fig.~\ref{GTP}. When $P<P_{c}$, BH's $G-T$ curve resembles a swallowtail diagram, indicating the phase transition. However, it cannot be concluded that phase transitions occur only when $P<P_{c}$. As previously analyzed, phase transitions can still occur even for $P\geq P_{c}$, where the phase transition point is characterized by the condition $\partial G/\partial T=\infty$, or equivalently, $C_{p}=\infty$. Moreover, due to the corrected first law, the swallowtail structure of the BH deviates from that of the Schwarzschild-AdS BH or the Van der Waals fluid. Only when $P$ is sufficiently small, or equivalently when $\xi$ is small enough, the $G-T$ curve of the BH closely resemble the swallowtail behavior observed in the Schwarzschild-AdS case.
\begin{figure}[htbp]
	\centering
	\includegraphics[width=1\textwidth]{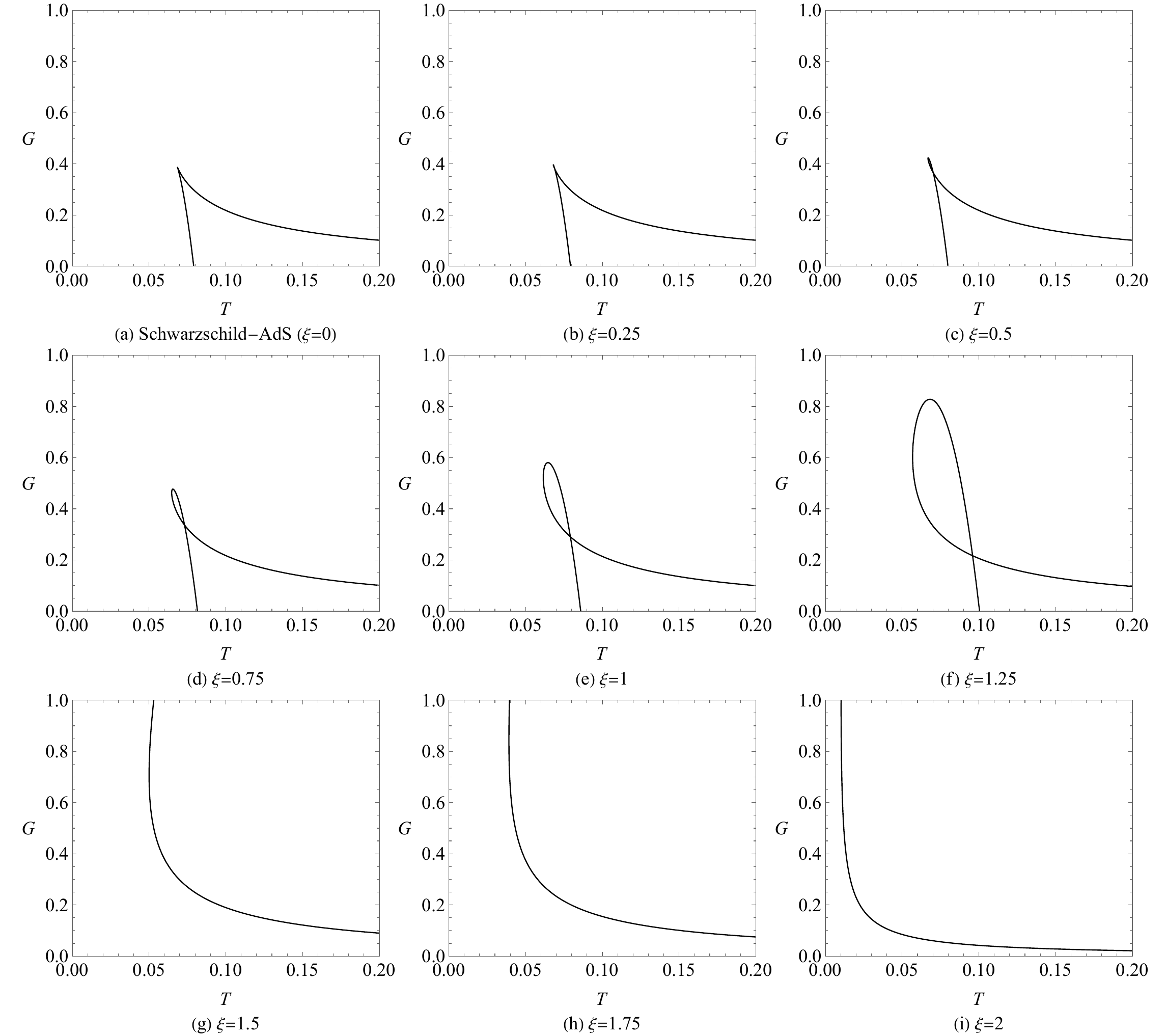}
	\caption{Gibbs free energy of the BH for different quantum parameters $\xi$. We set $P=5/216\pi$ (the critical pressure for $\xi=1.5$).}\label{GTxi}
\end{figure}

Fig.~\ref{GTxi} displays the Gibbs free energy of the BH at a fixed pressure for various values of the quantum parameter $\xi$. It is evident that when $\xi=0$, the $G-T$ curve of the BH reverts to that of the Schwarzschild-AdS case. As $\xi$ increases, the $G-T$ curve near the phase transition point becomes progressively smoother. Moreover, when $\xi$ is further increased such that the pressure exceeds the critical pressure, the $G-T$ curve of the BH deviates significantly from the Schwarzschild-AdS scenario.

\section{Joule-Thomson expansion}\label{Sect6}
The Joule-Thomson process of BHs essentially refers to their constant mass expansion process $dM=0$. In many AdS BHs, the Joule-Thomson process exhibits striking similarities: during the constant mass expansion, the pressure continuously decreases while the temperature changes accordingly. The inversion curve partitions the $T-P$ plane into distinct cooling and heating regions, and as the BH mass increases, the BH traverses increasingly broader ranges of pressure and temperature~\cite{JTcharged,JTBD1,JTBD2,NCthermo2,NCthermo4,FL3,dimension3,dimension4,JT1,JT2,JT3,JT4,JT5,JT6}.

\begin{figure}[htbp]
	\centering
	\begin{minipage}{0.49\textwidth}
		\centering
		\includegraphics[width=\linewidth]{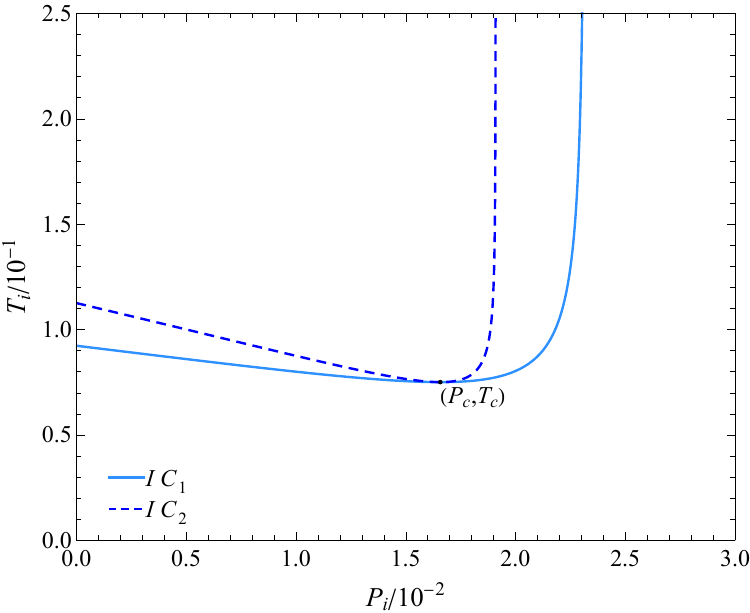}
		\caption{BH's inversion curves. We set $\xi=1$.}\label{inv}
	\end{minipage}
	\hfill
	\begin{minipage}{0.49\textwidth}
		\centering
		\includegraphics[width=\linewidth]{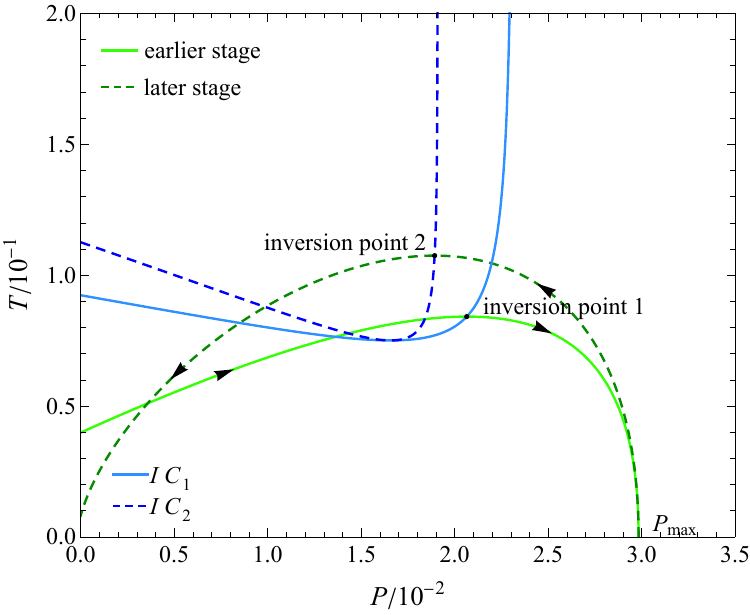}
		\caption{Joule-Thomson expansion of the BH. We set $\xi=1$, $M=5$.}\label{JTexample}
	\end{minipage}
\end{figure}

For the convenience of studying the constant mass expansion of the BH, we first rewrite the BH temperature and pressure in terms of the horizon radius and BH mass as parameters
\begin{align}
	P=-\frac{3\left(r_{+}-2M\right)\left(r_{+}^{3}-2M\xi^{2}+r_{+}\xi^{2}\right)}{8\pi r_{+}^{6}},\qquad\qquad\qquad\quad\label{eq6_1}\\
	T=\frac{-3\left(r_{+}^{3}-4M\xi^{2}+2r_{+}\xi^{2}\right)^{2}+\left(3r_{+}^{3}+2r_{+}\xi^{2}\right)\left|r_{+}^{3}-4M\xi^{2}+2r_{+}\xi^{2}\right|}{8\pi r_{+}^{5}\xi^{2}}.\label{eq6_2}
\end{align}
The Joule-Thomson coefficient is
\begin{equation}\label{eq6_3}
	\mu_{JT}=\left(\frac{\partial T}{\partial P}\right)_{M}=\frac{r_{+}\left(3r_{+}^{6}-8r_{+}^{3}\left(r_{+}-3M\right)\xi^{2}+4\left(8M-3r_{+}\right)r_{+}\xi^{4}-3\left(r_{+}^{3}+20M\xi^{2}-6r_{+}\xi^{2}\right)\left|r_{+}^{3}+2\left(r_{+}-2M\right)\xi^{2}\right|\right)}{6\left(r_{+}-3M\right)\xi^{2}\left|\left(r_{+}^{3}-4M\xi^{2}+2r_{+}\xi^{2}\right)\right|}.
\end{equation}
By setting $\mu_{JT}=0$, one gets the inversion mass $M_{i}$
\begin{align}
	M_{i}^{1}=\frac{9r_{i}^{3}+28r_{i}\xi^{2}+\sqrt{171r_{i}^{6}+204r_{i}^{4}\xi^{2}+64r_{i}^{2}\xi^{4}}}{60\xi^{2}},\quad r_{i}>\frac{\sqrt{2+\sqrt{22}}}{3}\xi,\label{eq6_4}\\
	M_{i}^{2}=\frac{3r_{i}^{3}+20r_{i}\xi^{2}+\sqrt{9r_{i}^{6}+60r_{i}^{4}\xi^{2}+40r_{i}^{2}\xi^{4}}}{60\xi^{2}},\quad r_{i}>\frac{\sqrt{2}}{2}\xi,\qquad\label{eq6_5}
\end{align}
where $r_{i}$ is the inversion radius. Thus, the inversion curve of the BH is divided into two branches (Fig.~\ref{inv}): for a given inversion radius, $M_{i}^{1}$ induces the first branch ($IC_{1}$), while $M_{i}^{2}$ induces the second branch ($IC_{2}$).
$IC_{1}$ and $IC_{2}$ are exactly tangent at their minimum points $T_{i}^{min}$, and aside from the tangency point, $T_{i}^{1}<T_{i}^{2}$ is identically satisfied. The phenomenon of multiple branches emerging in the inversion curve has also been observed in other types of BHs. For AdS BHs in quasitopological electromagnetism, the inversion curve may exhibit up to three branches~\cite{JoseJT}. Moreover, the inversion pressure and inversion temperature at this tangency point exactly coincide with the BH's critical pressure and critical temperature. That is $T_{i}^{min}=T_{c}$.

In the limit as $r_{+}\to\infty$, $IC_{1}$ and $IC_{2}$ both exhibit a maximum inversion pressure
\begin{equation}\label{eq6_6}
	P_{i}^{1max}=\frac{3+6\sqrt{19}}{400\pi\xi^{2}},\quad P_{i}^{2max}=\frac{3}{50\pi\xi^{2}}.
\end{equation}
This results in two constants that are independent of the quantum parameter $\xi$
\begin{equation}\label{eq6_7}
	\frac{P_{i}^{1max}}{P_{c}}=\frac{6}{125}\left(3+6\sqrt{19}\right),\quad\frac{P_{i}^{2max}}{P_{c}}=\frac{144}{125}.
\end{equation}
Fig.~\ref{JTexample} depicts the constant-mass expansion process of the BH. In the figure, the arrows indicate the direction of increasing horizon radius $r_{+}$, representing the expansion process of the BH. It is evident that the constant-mass expansion of the BH can be divided into two distinct stages. In the earlier stage, the pressure increases until reaching the BH's maximum pressure $P_{max}$; in the later stage, the pressure decreases. The black dots in the figure denote the inversion points experienced by the BH in each stage.

It should be noted that, because the BH pressure increases rather than decreases during the pre-expansion stage, using the sign of $\mu_{JT}$ to differentiate between cooling and heating processes is no longer appropriate. More accurately, the sign in $\left(\partial T/\partial V\right)_{M}$ (or equivalently, $\left(\partial T/\partial r_{+}\right)_{M}$) should be employed to distinguish between the cooling process and the heating process.
\begin{equation}\label{eq6_8}
\left(\frac{\partial T}{\partial V}\right)_{M}=
\begin{cases}
	<0, & \text{cooling point},\\
	=0, & \text{inversion point},\\
	>0, & \text{heating point}.
\end{cases}
\end{equation}
Thus, as depicted in Fig.~\ref{JTexample}, during both the earlier stage and the later stage, the BH initially undergoes a heating phase before transitioning into a cooling phase at the inversion point.

Moreover, considering that for a given inversion radius $r_{i}$, $M_{i}^{1}>M_{i}^{2}$ always holds, and $M_{i}^{1}\left(r_{i}\right)$, $M_{i}^{2}\left(r_{i}\right)$ are both increasing functions, for a constant mass process, the BH necessarily first passes through the inversion point on $IC_{1}$, followed by that on $IC_{2}$. In other words, $IC_{1}$ precisely corresponds to the inversion points experienced by BHs in the earlier stage, whereas $IC_{2}$ pertains to the inversion points attainable by BHs in the later stage. In addition, since a given point in the $\left(P,T\right)$ plane may correspond to several distinct BH phases, the inversion curve does not, by itself, partition the $\left(P,T\right)$ plane into separate cooling and heating regions. In other words, in a constant- ass process, the intersection of the expansion curve and inversion curve is not necessarily the inversion point. This peculiar feature is also similar to that of the BH discussed in Ref.~\cite{JoseJT}.

The earlier and later stages of the BH are connected at the point of maximum pressure $P_{max}$, at which the horizon radius $r_{+}$ is
\begin{equation}\label{eq6_9}
	r_{+c}^{JT}=\frac{6^{\frac{2}{3}}\xi^{\frac{4}{3}}-6^{\frac{1}{3}}\xi^{\frac{2}{3}}\gamma^{\frac{2}{3}}}{3\gamma^{\frac{1}{3}}},
\end{equation}
where $\gamma=\sqrt{81M^{2}+6\xi^{2}}-9M$.
\begin{figure}[htbp]
	\centering
	\begin{minipage}{0.49\textwidth}
		\centering
		\includegraphics[width=\linewidth]{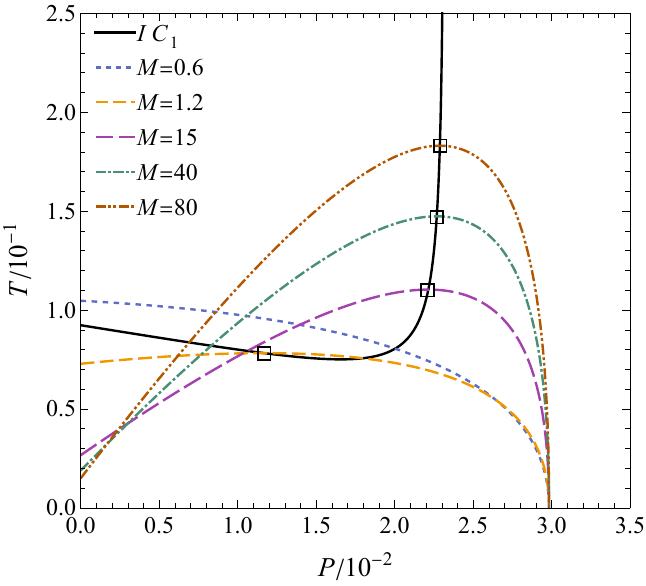}
		\caption{Earlier stage of BH's constant mass expansion. The points circled by small squares in the figure denote the inversion points. We set $\xi=1$.}\label{JTe}
	\end{minipage}
	\hfill
	\begin{minipage}{0.49\textwidth}
		\centering
		\includegraphics[width=\linewidth]{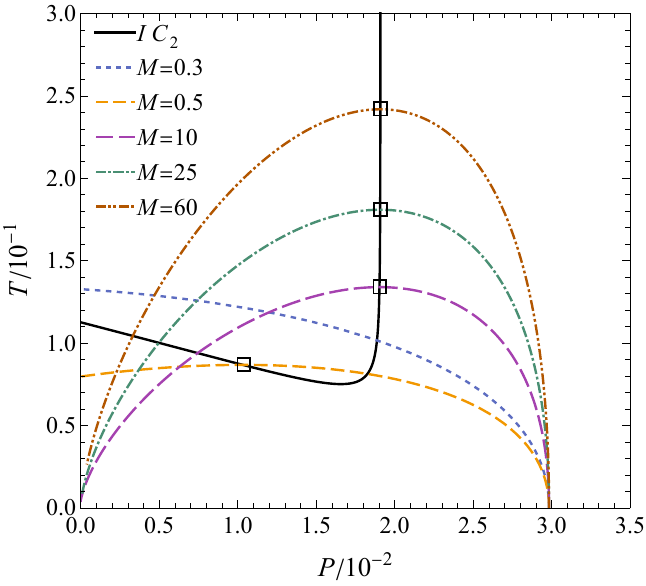}
		\caption{Later stage of BH's constant mass expansion. The points circled by small squares in the figure denote the inversion points. We set $\xi=1$.}\label{JTl}
	\end{minipage}
\end{figure}

Fig.~\ref{JTe} and Fig.~\ref{JTl} respectively depict the $T-P$ diagrams for BHs with varying masses during the earlier stage and later stage of expansion. It is evident that, for sufficiently small BH masses, no inversion point exists. By setting $P_{i}=0$, the minimum inversion mass $M_{i}^{min}$ for each case can be determined
\begin{equation}\label{eq6_10}
	M_{ie}^{min}=\frac{\sqrt{770+187\sqrt{22}}}{54}\xi,\quad M_{il}^{min}=\frac{1}{2\sqrt{2}}\xi.
\end{equation}
The minimum inversion mass also could be derived from Eq.~\ref{eq6_4} and Eq.~\ref{eq6_5} directly.
\begin{equation}\label{eq6_11}
	M_{ie}^{min}=M_{i}^{1}\left(\frac{\sqrt{2+\sqrt{22}}}{3}\xi\right),\quad M_{il}^{min}=M_{i}^{2}\left(\frac{\sqrt{2}}{2}\xi\right).
\end{equation}
In Fig.~\ref{JTe} and Fig.~\ref{JTl}, the first constant mass curve in each does not exhibit an inversion point, which is because
\begin{equation}\label{eq6_12}
	0.6<M_{ie}^{min}\approx0.75,\quad0.3<M_{il}^{min}\approx0.35.
\end{equation}
\begin{figure}[H]
	\centering
	\begin{minipage}{0.49\textwidth}
		\centering
		\includegraphics[width=\linewidth]{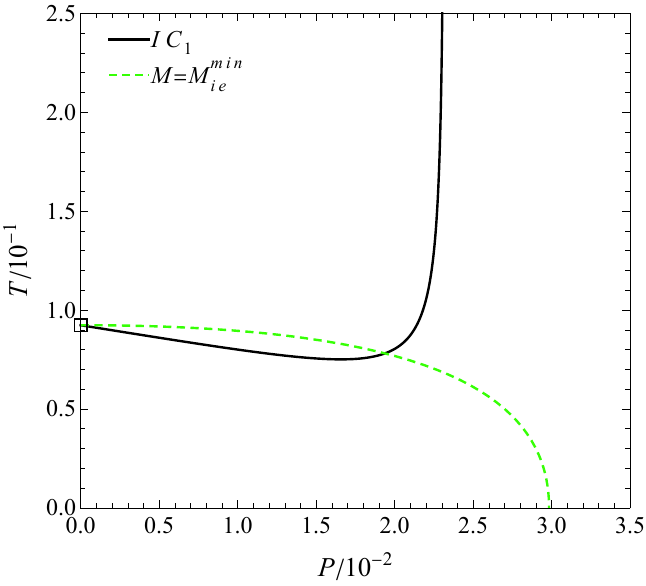}
		\caption{Earlier stage of constant mass expansion for BH with $M_{ie}^{min}$. The points circled by small squares in the figure denote the inversion points. We set $\xi=1$.}\label{qqljt}
	\end{minipage}
	\hfill
	\begin{minipage}{0.49\textwidth}
		\centering
		\includegraphics[width=\linewidth]{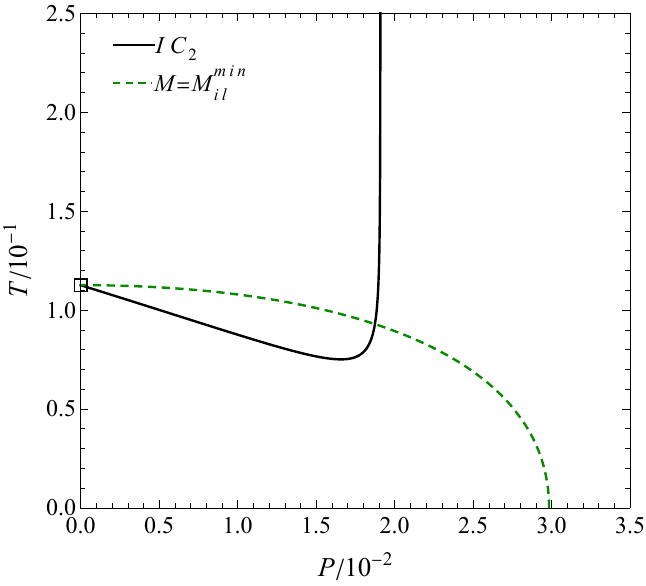}
		\caption{Later stage of constant mass expansion for BH with $M_{il}^{min}$. The points circled by small squares in the figure denote the inversion points. We set $\xi=1$.}\label{hqljt}
	\end{minipage}
\end{figure}
As seen in Fig.~\ref{qqljt} and Fig.~\ref{hqljt}, the BHs with $M_{i}^{min}$ reach their critical points at $P=0$.

\section{Conclusion and outlook}\label{Sect7}
In this article, we investigated the thermodynamics in the extended phase space of a quantum corrected BH proposed recently. Our analysis demonstrates that the phase transition behavior of the BH is analogous to that of conventional Schwarzschild-AdS BH. In particular, there exists a critical temperature $T_{c}$ such that for any BH with a temperature exceeding $T_{c}$, the small BH phase and the large BH phase become separated, and no phase transition occurs. Due to the effect of the quantum parameter $\xi$, the BH equation of state $P\left(T,r_{+}\right)$ bifurcates into two branches with completely opposite stability properties. One branch converges to the Schwarzschild-AdS BH as $\xi\to0$, and it is associated with a phase transition pressure lower than the critical pressure $P_{c}$, whereas the other branch exhibits a phase transition pressure greater than $P_{c}$, which is not observed in conventional Schwarzschild-AdS BHs.

The temperature and heat capacity $C_{p}$ of the BH closely resemble those of the Schwarzschild-AdS BH. Specifically, the small BH phase exhibits a negative heat capacity, indicative of an unstable configuration, whereas the large BH phase shows a positive heat capacity, corresponding to a stable state. Moreover, the BH does not exhibit any region of negative temperature. Moreover, owing to the corrected first law, the Gibbs free energy curve $G-T$ in the vicinity of the BH phase transition point becomes smooth, thereby deviating from the behavior observed in conventional Schwarzschild-AdS BHs.

In addition, we conducted a detailed study of the Joule-Thomson expansion process of the BH. Our analysis reveals that the constant mass expansion process of the BH has two stages. In the earlier stage, the BH pressure increases with expansion until it reaches its maximum value; subsequently, during the later stage, the pressure gradually decreases. Consequently, the inversion curve of the BH splits into two branches, corresponding to the inversion points that may be experienced in the earlier and later stages, respectively. Moreover, each expansion stage is associated with a minimum inversion mass $M_{i}^{min}$, below which any BH undergoing a constant mass expansion process (in either stage) will not exhibit an inversion point.

It is our hope that these findings could offer novel insights into the interplay between quantum corrections and classical BH thermodynamics, and serve as a valuable reference for future research.

In this work, our study is based on the assumption that the quantum gravity theory responsible for generating this quantum corrected BH decouples from the cosmological constant. This assumption, also adopted in Ref.~\cite{FL3}, is generally regarded as valid since the cosmological constant, being a constant, typically does not interact with other fields. Nevertheless, some modified gravity models suggest that the cosmological constant may be linked to other interactions. For instance, in Ref.~\cite{BTZ}, the authors constructed a gravastar in non-commutative BTZ geometry, finding the non-commutativity parameter plays a role of the cosmological constant for gravastar formation and stability. If this theory is correct, the cosmological constant might be suggested as a non-commutative effect. In the context of Lorentz symmetry breaking theories, researchers have found that the choice of the self-interaction potential in the Kalb-Ramond field differs between the cases of a zero and a nonzero cosmological constant~\cite{KR}. It is foreseeable that the cosmological constant may no longer be merely a constant term, but instead could couple with a variety of modified gravity models. In addition, although the quantum parameter $\xi$ is presumed to be extremely small, its precise value remains undetermined. One approach is to estimate $\xi$ by comparing the observed BH shadow radius with the theoretical value predicted for a Schwarzschild BH~\cite{xioptics,xivalue1,xivalue2}. However, this method can only constrain $\xi$ to be on the order of the Schwarzschild radius, which is evidently much larger than the Planck length $\ell_{p}$. Therefore, seeking alternative methods to constrain this quantum parameter more precisely represents an interesting and worthwhile topic for future research.

Moreover, unlike most other AdS BHs, the phase structure of the BH studied in this work exhibits significant deviations from that of the Van der Waals system. It remains an important issue for future investigation to uncover the deeper physical mechanisms responsible for this phenomenon, as well as to explore whether other different quantum effects could result in similar thermodynamic properties.

\section*{Conflicts of interest}
The authors declare that there are no conflicts of interest regarding the publication of this paper.

\section*{Acknowledgments}
We want to thank School of Physical Science and Technology, Lanzhou University.

\section*{Data availability}
This work is a theoretical study. No data was used or generated during this research.

\bibliography{paper}

\end{document}